\documentclass[twocolumn,aps,prl,showpacs,superscriptaddress]{revtex4-1}
\usepackage{epsfig}
\usepackage{graphics}
\usepackage{amssymb}
\usepackage{amsmath}
\usepackage{epstopdf}
\usepackage{subeqnarray}
\usepackage{setspace}
\DeclareGraphicsExtensions{.pdf,.eps,.png,.jpg,.mps}

\usepackage{epsfig}
\usepackage{siunitx}

\begin{document}

\title{Photonic Nonlinearities via Quantum Zeno Blockade}

\author{Yu-Zhu Sun}
\affiliation{Center for Photonic Communication and Computing and Department of Physics and Astronomy, Northwestern University, 2145 Sheridan Road, Evanston, Illinois 60208-3112, USA}
\author{Yu-Ping Huang}
\email{yphuangpx@gmail.com}
\affiliation{Center for Photonic Communication and Computing and Department of Electrical Engineering and Computer Science, Northwestern University, 2145 Sheridan Road, Evanston, Illinois, 60208-3118, USA}
\author{Prem Kumar}
\affiliation{Center for Photonic Communication and Computing and Department of Physics and Astronomy, Northwestern University, 2145 Sheridan Road, Evanston, Illinois 60208-3112, USA}
\affiliation{Center for Photonic Communication and Computing and Department of Electrical Engineering and Computer Science, Northwestern University, 2145 Sheridan Road, Evanston, Illinois, 60208-3118, USA}

\date{\today}

\begin{abstract}
Realizing optical-nonlinear effects at a single-photon level is a highly desirable but also extremely challenging task, because of both fundamental and practical difficulties. We present an avenue to surmounting these difficulties by exploiting quantum Zeno blockade in nonlinear optical systems. Considering specifically a lithium-niobate microresonator, we find that a deterministic phase gate can be realized between single photons with near-unity fidelity. Supported by established techniques for fabricating and operating such devices, our approach can provide an enabling tool for all-optical applications in both classical and quantum domains.
\end{abstract}
\maketitle

At the very heart of classical and quantum optics lie phenomena resulting from optical nonlinearities. First observed more than half a century ago \cite{Franken61}, such phenomena have since become the foundation for interdisciplinary applications, such as squeezed light sources \cite{kimblesqueeze}, biological microscopy \cite{GuoOL97}, and entanglement generation \cite{Horodecki09}. Recently, the quest for information processing by all-optical means has fueled new studies of optical phenomena in an extreme quantum regime involving only a few photons \cite{OptTransRevNature10,NieChu00}. This requires optical nonlinearities that are orders of magnitude higher than those achievable with existing optical media \cite{SFGQKD11,CPC_Nature2011}. Although this drawback can be overcome by combining strong cavity enhancement with resonant coupling between photons and (effective) atoms  \cite{PhotonBlockade05,VolzNP12}, the implementation requires large setups and operation in near-zero-temperature environment, making such systems unsuitable for practical use. In contrast, schemes based on post selection \cite{KniLafMil01} or feed forward \cite{RauBri01} can be implemented with only linear-optics instruments. Such schemes, however, are inherently probabilistic and thus their use is hard to justify in large-scale applications.

Highly off-resonant optical nonlinearities, on the other hand, do not suffer from the aforementioned issues and are thus potentially viable for photonic information processing tasks on a large scale. It was first pointed out in \cite{ChuangYamamoto95} and later followed by many others \cite{LukImaPRL00,WangPRL06} that intense cross-phase modulation (XPM) in Kerr-nonlinear media could produce a deterministic phase gate between single photons. The underlying intuition is straightforward: since a classical beam containing many photons can produce a large XPM phase shift with a weak Kerr nonlinearity, a quantum beam containing only a few or even a single photon should be able to create similar XPM phase shift, provided the nonlinearity is sufficiently giant. This logic, unfortunately, is based on an incorrect single-mode argument. By taking into account the inherent multimode nature of light propagation in a Kerr medium, it was recently discovered that no useful XPM effect can be produced in such systems even with an unrealistically giant Kerr nonlinearity \cite{ShapiroPRA06,BanaclochePRA10}. The fundamental reason turns out to be that causality prohibits XPM phase shift of any non-negligible amount without significant quantum noise.

It remains an outstanding challenge---not only because of implementation difficulties but also due to the fundamental restrictions---to construct practical nonlinear optical devices suitable for operation at the single photon level. In this Letter, we propose to surmount this challenge by employing quantum Zeno blockade (QZB), which occurs when a quantum system interacts with external degrees of freedom through a nonlinear channel. When the nonlinear coupling is strong, occupation of a certain quantum mode by a single photon can ``block'' (more precisely, suppress) additional photons from coupling into that mode \cite{MisSud77,HuaMoo08}. In a nutshell, the nonlinear interaction functions as a continuous measurement to freeze the coupling dynamics---a quantum Zeno effect. The interaction can be dissipative, like that of two-photon absorption \cite{ZenoGate04,HuangKumar02PRL}, or be coherent such as sum or difference frequency generation \cite{HuaAltKum10-2}. Using QZB, quantum optical Fredkin gates can be implemented in an ``interaction-free'' manner \cite{HuangKumar10,HuaKumSwitch11}. Here we employ QZB to realize strong nonlinear effects between single photons. Specifically by considering a $\chi^{(2)}$ system of a prism-coupled lithium-niobate (LN) microdisk resonator, we show that strong XPM effects can be produced between single photons under realizable parameter settings. When the input single photons are in the form of Gaussian pulses, they become entangled at the output. However, when the input photons are prepared in exponential waveforms that are time-reversed replicas of the cavity leakage modes \cite{kimblerev}, a deterministic phase gate can be realized. This result highlights a potentially enabling pathway for implementing practical photonic information processing. Our approach is generally extendable to a variety of nonlinear optical systems of traveling-wave or resonator designs.

\begin{figure}
 \epsfig{figure=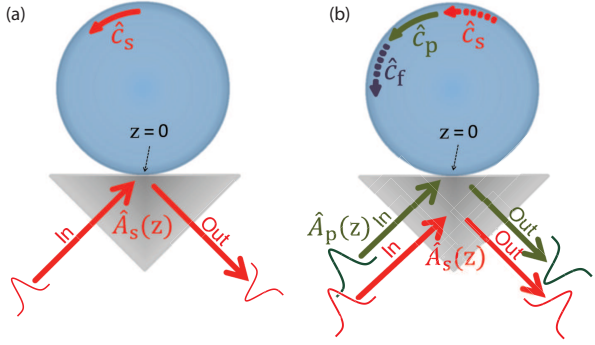,width=8cm}
      \caption{A schematic of the gate setup. $\hat{A}_{\mathrm{s (p)}}$ is the annihilation operator for the propagating signal (pump) field. $\hat{c}_{\mathrm{s (p,f)}}$ is the annihilation operator for the intracavity signal (pump, sum-frequency) field. Dashed arrows indicate the interaction-free nature of the gate's operation. (a) Pump OFF: The signal exits with a $\pi$ phase shift. (b) Pump ON: The signal exits with its phase unchanged. The pump output gains a $\pi$ shift.\label{fig1}}
\end{figure}

In Fig.~\ref{fig1}, we schematize the operation of our phase gate. It consists of a LN microdisk cavity evanescently coupled to a prism. The cavity is designed to be in resonance with both the pump (control) and the signal (target). The $\chi^{(2)}$ nonlinearity of LN can lead to intracavity sum-frequency generation (SFG) or difference-frequency generation (DFG) between the signal and the pump. Only SFG is considered in the following as a specific example. Figure~\ref{fig1}(a) shows the pump-OFF case. Through the prism, the signal field evanescently couples into the disk and then exits with a $\pi$ phase shift relative to its input. Figure~\ref{fig1}(b) shows the pump-ON case. The pump field, applied ahead of the signal, couples in and then out of the disk with a $\pi$ phase shift. When the signal photon arrives, the presence of the intracavity pump field and the large nonlinear coupling strength provide high potential for SFG. Depending on the lifetime of the sum-frequency (SF) field in the cavity, gate operation falls into two regimes: coherent quantum zeno (CQZ) and incoherent quantum zeno (IQZ) \cite{HuaAltKum10-2}. In the CQZ case, the cavity is resonant for the SF field with high intrinsic quality factor $Q_\mathrm{f}^\mathrm{i}$ ($\geq Q_\mathrm{s,p}^\mathrm{i}$). Coherent SFG process thus works as a continuous measurement, which shifts the disk away from resonance with the signal, preventing the signal photon from entering the cavity. In the IQZ case, a short-lived cavity SF field, together with the high potential for SFG, opens a strong dissipation channel for the intracavity signal field, destroying the cavity resonance and preventing the signal photon from entering the cavity. In either the CQZ or the IQZ regime, the signal's coupling into disk is thus suppressed and it reflects (exits) with its phase unchanged. Below we consider the CQZ case in detail; see the supplementary material for the IQZ case.

\begin{figure}
 \epsfig{figure=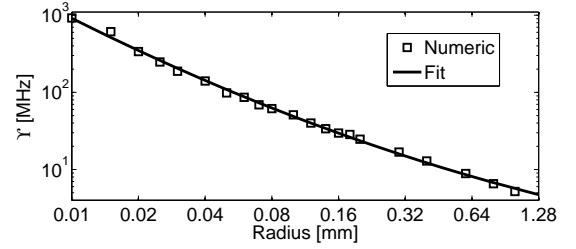,width=8cm}
      \caption{$\Upsilon$ versus $R$ on a $\log$ scale. The plotted curve is $\log_{10}\Upsilon=0.073(\log_{10} R)^2-0.77\log_{10} R +1.74$, which is obtained by least-square fitting.\label{fig2}}
\end{figure}
The eigenmodes of the microdisk cavity are whispering gallery modes (WGMs). See details in the supplementary material \cite{sm}. 
For a typical LN microdisk with radius $R\simeq1.5$ mm and $Q^\mathrm{i}> 10^{7}$ \cite{IIcSavMat04}, the resonance linewidth is much smaller than the cavity's free-spectral range. It is thus valid to consider only a single cavity mode being excited by the corresponding applied external quasimonochromatic field. We then denote the cavity-mode annihilation operators as $\hat{c}_{\mu}$ with $\mu \in \{\mathrm{s,p,f}\}$ representing the signal, pump, and SF fields, respectively. They satisfy $[\hat{c}_{\mu},\hat{c}^{\dagger}_{\mu'}]=\delta_{\mu\mu'}$ and $[\hat{c}_{\mu},\hat{c}_{\mu'}]=0$. The corresponding external propogating-field annihilation operators are denoted by $\hat{A}_{\mu}$, which satisfy $[\hat{A}_{\mu}(z),\hat{A}^{\dagger}_{\mu'}(z')]=\delta_{\mu\mu'}\delta(z-z')$ and $[\hat{A}_{\mu}(z),\hat{A}_{\mu'}(z')]=0$.

To capture the temporal behavior of pulsed optical signals, a temporally multimode quantum description is needed. We consider an effective real-space Hamiltonian specific to our gate system, as follows \cite{Shen2009}:
\begin{equation}
\label{Hamiltonian}
    \begin{aligned}
        \hat{H}= &\sum_{\mu=\mathrm{s,p,f}} \bigg[ \hbar\omega_{\mu}\hat{c}_{\mu}^{\dagger}\hat{c}_{\mu}
        +\int \mathrm{d}z\, \hat{A}_{\mu}^{\dagger}(z)\hbar(\omega_{\mu}-iv_\mathrm{g}\frac{\partial}{\partial z})\hat{A}_{\mu}(z)\\
        &\phantom{\sum_{\mu=s,p,f}}+\hbar \sqrt{v_\mathrm{g}\frac{\omega_{\mu}}{Q_{\mu}^{\mathrm{c}}}}\int \mathrm{d}z\,\delta(z)
        (\hat{A}_{\mu}^{\dagger}(z)\hat{c}_{\mu}+\hat{A}_{\mu}(z)\hat{c}_{\mu}^{\dagger})\bigg]\\
        &+\hbar\left(\Upsilon \hat{c}_\mathrm{s}\hat{c}_\mathrm{p}\hat{c}_\mathrm{f}^{\dagger}+
        \Upsilon ^{\ast}\hat{c}_\mathrm{s}^{\dagger}\hat{c}_\mathrm{p}^{\dagger}\hat{c}_\mathrm{f}\right),
    \end{aligned}
\end{equation}
where $\omega_{\mu}$ is the carrier frequency of the field $\hat{A}_{\mu}$ and $v_\mathrm{g}$ is the group velocity which is assumed to be the same for all the pulsed fields. Since the carrier-frequency terms in Eq.~(\ref{Hamiltonian}) do not influence the gate's dynamics, we ignore terms containing $\hbar \omega_{\mu}$ for simplicity. The third term in Eq.~(\ref{Hamiltonian}) describes the coupling between the external propagating fields and the cavity fields, where the coupling quality factor $Q_{\mu}^\mathrm{c}$ places a bandwidth limit on the input pulses that can be properly coupled-in without distortion. The last two terms in Eq.~(\ref{Hamiltonian}) describe the intracavity SFG process, where $\Upsilon$ is the nonlinear coupling coefficient. 
%
In Fig.~\ref{fig2} we plot $\Upsilon$ as a function of $R$ calculated using analytical WGM profiles (see supplementary material \cite{sm} for details), where $\Upsilon=140$ and $337$ MHz are shown to be obtained for $R\approx$ 50 and \SI{20}{\micro\metre}, respectively. Such large $\Upsilon$ values are potentially realizable because high-quality LN microdisks with $R$ as small as \SI{40}{\micro\metre} have been fabricated via manual polishing techniques \cite{Dmitry}. Even smaller disks are expected to be fabricated in the near future by adopting advanced automated machining and polishing.

In arriving at Eq.~(\ref{Hamiltonian}), we have assumed $Q^\mathrm{c}_\mu\ll Q^\mathrm{i}_\mu$ and thus have neglected terms that describe decay of the WGMs. $Q^\mathrm{i}_\mu$ is fundamentally limited by bending loss and light absorption within the disk. Because LN's large refractive index, the bending loss allows for $Q^\mathrm{i}_\mu > 10^{14}$ for disks with $R>$ \SI{15}{\micro\metre} \cite{HeeBonKal07,toroidal-q}. On the other hand, the absorption loss in commercially-available LN crystals is usually $\sim 10^{-4}$/cm or higher, which imposes a high limit of $Q^{\mathrm{i}}_\mu \sim 10^8$, as typically demonstrated \cite{JPL_MicrodiskPRL10}. By custom doping LN crystals, however, ultralow absorption loss ($\lesssim 10^{-5}$ $\mathrm{cm} ^{-1}$, bounded by measurement precision) has been demonstrated over a wide spectral range (800--2000 nm) \cite{absorpLN}. Using such crystals, microdisks with $Q^\mathrm{i}_\mu \gtrsim 10^9$ can be fabricated. Note that in practice $Q^\mathrm{i}_\mu$ can also be limited by roughness of the disk's surface. This effect, however, was not found to be significant, even for a manually-polished disk with $R=$ \SI{40}{\micro\metre} \cite{Dmitry}. Based upon the above analysis, we will take $Q^\mathrm{i}_\mu=10^9$, for which decay of the WGMs can be ignored when $Q^{\mathrm{c}}_\mu \lesssim 10^8$. For the setup depicted in Fig.~\ref{fig1}, the latter conditions can be realized by adjusting the distance between the disk and the prism.

\begin{figure}
 \epsfig{figure=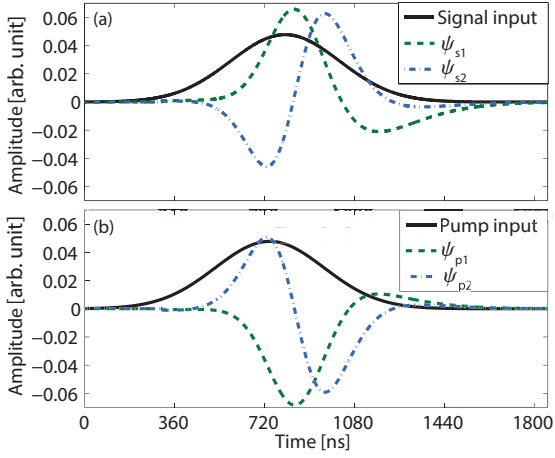,width=8cm}
      \caption{Gate performance with Gaussian pump and signal pulses. (a) Signal input and the first two Schmidt eigenmodes for the signal output. (b) Pump input and the first two Schmidt eigenmodes for the pump output. \label{fig3}}
\end{figure}
In our gate system, the joint quantum state of the signal, the pump, and the SF photons can be written in its most general form as: $
|\Psi(t)\rangle=\Big[\mathop{\int \!\!\! \int} \mathrm{d}z \mathrm{d}z' \phi_\mathrm{sp}(z,z',t)\hat{A}_\mathrm{s}^{\dagger}(z) \hat{A}_\mathrm{p}^{\dagger}(z')
+\int \mathrm{d}z \,\phi_\mathrm{f}(z,t)\hat{A}_\mathrm{f}^{\dagger}(z)
+\int \mathrm{d}z\, \phi_\mathrm{s}(z,t)\hat{A}_\mathrm{s}^{\dagger}(z)\hat{c}_\mathrm{p}^{\dagger}
+\int \mathrm{d}z \,\phi_\mathrm{p}(z,t)\hat{A}_\mathrm{p}^{\dagger}(z)\hat{c}_\mathrm{s}^{\dagger}+e_\mathrm{sp}(t)\hat{c}_\mathrm{s}^{\dagger}\hat{c}_\mathrm{p}^{\dagger}
+e_\mathrm{f}(t)\hat{c}_\mathrm{f}^{\dagger}\Big]|0\rangle$, where $|0\rangle$ is the vacuum state. The first term represents the state in which both the signal and the pump photons are in external traveling-wave modes with $\phi_\mathrm{sp}(z,z',t)$ as their joint wavefunction. Similarly, in the second term $\phi_\mathrm{f}(z,t)$ is the wavefunction for the external SF field. The third and fourth terms, in contrast, represent quantum states with one photon in the propagating mode and one in the cavity mode with $\phi_\mathrm{s}$ ($\phi_\mathrm{p}$) as the product of the signal (pump) wave function and the cavity excitation amplitude for the pump (signal). Finally, $e_\mathrm{f}$ ($e_\mathrm{sp}$) is the SF (product of signal and pump) cavity-excitation amplitude(s).

Plugging the Hamiltonian of Eq.~(\ref{Hamiltonian}) and the quantum state $|\Psi(t)\rangle$ into the Schr\"odinger equation, the following partial differential equations can be derived:
\begin{subeqnarray}
    \label{PDE}
    \slabel{PDE1}
    \partial_{t}\phi_\mathrm{f}&=&-v_\mathrm{g}\partial_{z}\phi_\mathrm{f}-i\Omega_\mathrm{f}\delta(z)e_\mathrm{f},\\
    \slabel{PDE2}
    \partial_{t}\phi_\mathrm{sp}&=&-v_\mathrm{g}(\partial_{z}+\partial_{z'})\phi_\mathrm{sp}-i\sum_{\nu}\Omega_{\nu}\Lambda_{\nu'},\\
    \slabel{PDE3}
    \partial_{t}\phi_{\nu}&=&-v_\mathrm{g}\partial_{z}\phi_{\nu}-i\Omega_{\nu}\delta(z)e_\mathrm{sp}-i\Omega_{\nu'}\Gamma_{0\nu},\\
    \slabel{PDE4}
    \partial_{t}e_\mathrm{sp}&=&-i\sum_{\nu}\Omega_{\nu}\phi_{\nu}(0)-i\Upsilon^{\ast}e_\mathrm{f},\\
    \slabel{PDE5}
    \partial_{t}e_\mathrm{f}&=&-i\Omega_\mathrm{f}\phi_\mathrm{f}(0)-i\Upsilon_\mathrm{sp}e_\mathrm{sp},
\end{subeqnarray}
where $\Omega_{\mu}=(v_\mathrm{g}\omega_{\mu}/Q_{\mu}^{\mathrm{c}})^{1/2}$ and $\nu,\nu' \in \{\mathrm{s,p}\}$, $\nu\neq\nu'$. In Eq.~(\ref{PDE2}), $\Lambda_\mathrm{s}=\delta(z')\phi_\mathrm{s}(z,t)$ and $\Lambda_\mathrm{p}=\delta (z)\phi_\mathrm{p}(z',t)$. In Eq.~(\ref{PDE3}), $\Gamma_{0\mathrm{s}}=\phi_\mathrm{sp}(z,0,t)$ and $\Gamma_{0\mathrm{p}}=\phi_\mathrm{sp}(0,z',t)$. In Eq.~(\ref{PDE}), the output joint wavefunction for the signal and pump photons can be decomposed as $\phi_\mathrm{sp}(z,z',t)=\sum^\infty_{n=1} a_{n}\tilde{\psi}_{\mathrm{s}n}(z,t)\tilde{\psi}_{\mathrm{p}n}(z',t)$ ($z,z'>0$), where $\tilde{\psi}_{\mathrm{s}n}$ and $\tilde{\psi}_{\mathrm{p}n}$ are pair-wise eigenfunctions for the output modes of the signal and pump photons, respectively, with eignvalues $a_n$, which are ordered such that $a_1>a_2>\cdot\cdot\cdot\ge 0$. The corresponding two-photon output state can then be written as $|\Psi_\mathrm{out}\rangle=\sum^\infty_{n=1} a_n |nn\rangle$, where $|nn\rangle=\mathop{\int \!\!\! \int} dt\,dt' \psi_{\mathrm{s}n}(t)\psi_{\mathrm{p}n}(t')\hat{A}^\dag_\mathrm{s}(v_\mathrm{g}t)\hat{A}^\dag_\mathrm{p}(v_\mathrm{g}t')|0\rangle$ and $\psi_{{\mathrm{p}n},\mathrm{s}n}(t)=v_\mathrm{g}\tilde{\psi}_{{\mathrm{p}n},\mathrm{s}n}(z\rightarrow 0^+,t)$. A near-unity $a_1^2$ together with a large overlap between $\psi_{\mathrm{s}1}$ and the input signal $\psi_\mathrm{s}$, as quantified by the fidelity $|\langle\psi_\mathrm{s}|\psi_{\mathrm{s}1}\rangle|^{2}$, then signify high-performance switching when the pump is on.

\begin{figure}
\epsfig{figure=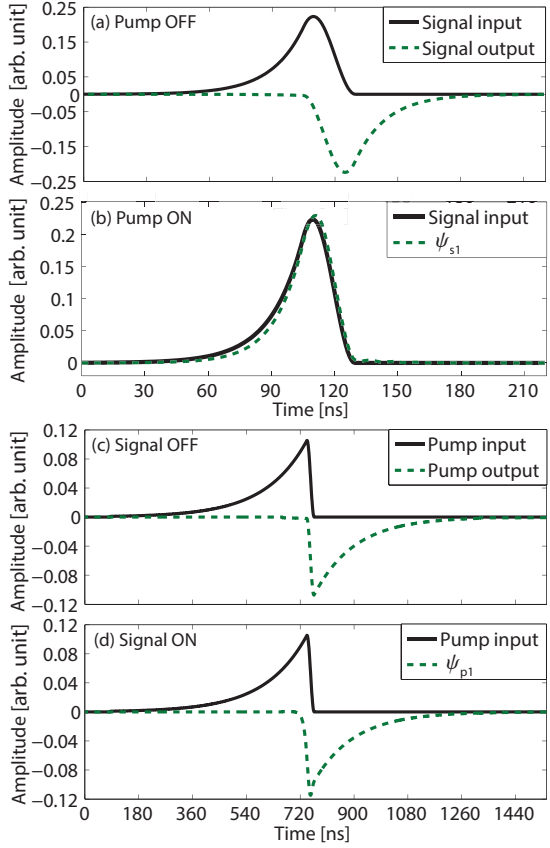,width=8cm}
	\caption{CQZ gate performance with the proposed pulse shapes. Input (output) is defined as the pulse after it has evolved the same period of time without (with) passing the gate. The temporal reference points are the same in all the plots. (a) Pump OFF: the signal output is phase shifted by $\pi$ and temporally reversed. (b) Pump ON: The output signal is well preserved. (c) Signal OFF: the pump behaves the same as the signal in the pump-OFF case. (d) Signal ON: the pump behaves as if there is no signal photon. \label{fig4}}
\end{figure}
To examine the gate performance, we first consider $Q^\mathrm{c}_\mathrm{s,p,f}=10^8$, $\Upsilon=610$ MHz, and two $500$ ns (FWHM) Gaussian signal and pump pulses with a separation of $60$ ns. The simulation results are shown in Fig.~\ref{fig3}, where the output photons turn out to be in an entangled two-photon state with $a_1$ = 0.77, $a_2$ = 0.52, $a_3$ = 0.25, and $a_{n>3}\simeq 0$. This result points to a viable approach for \emph{on-demand} generation of entanglement using two initially uncorrelated photons, which could find important applications in implementing deterministic quantum-information processing. The physics underlying this behavior is in some way similar to what was found during the study of single-photon cross-phase modulation in the fast-response regime \cite{ShapiroPRA06}. In a nutshell, it is because the QZB is effective only if the pump photon is in the cavity. Therefore, the phase of the signal photon can be switched only when it arrives within the cavity lifetime of the pump photon. In the above example, the cavity lifetime is about $100$ ns for both the signal and pump photons, which is five times smaller than their pulse widths. As a result, depending on the temporal location of the pump photon within its pulse duration, the output signal photon will be in a superposition of phase changed and unchanged sates. Because of the quantum uncertainty inherent in the pump-photon location, the two photons therefore exit the cavity in an entangled state. This phenomenon can be intuitively understood by considering a toy model in which the pump and the signal photons are initially in states $(|t_0\rangle_\mathrm{p}+|t_1\rangle_\mathrm{p})/\sqrt{2}$ and $(|t_0+\Delta \rangle_\mathrm{s}+|t_1+\Delta\rangle_\mathrm{s})/\sqrt{2}$, respectively, where $\{|t\rangle_\mathrm{p(s)}\}$ are orthonormal time modes centered at $t$ for the pump (signal). The times $t_1$ and $\Delta$ are chosen such that the signal photon in $|t_{0(1)}+\Delta\rangle_\mathrm{s}$ arrives at the cavity when the pump photon in $|t_{0(1)}\rangle_\mathrm{p}$ has already coupled into the cavity, but the photon in $|t_{1(0)}\rangle_\mathrm{p}$ has not (exited). Under this condition, the output pump and signal photons in the presence of an ideal QZB effect will be in an entangled state of $\big(|t_0\rangle_\mathrm{p}|t_1+\Delta\rangle_\mathrm{s}+|t_1\rangle_\mathrm{p}|t_0+\Delta\rangle_\mathrm{s}
-|t_0\rangle_\mathrm{p}|t_0+\Delta\rangle_\mathrm{s}-|t_1\rangle_\mathrm{p}|t_1+\Delta\rangle_\mathrm{s}\big)/2$.

\begin{figure}
 \epsfig{figure=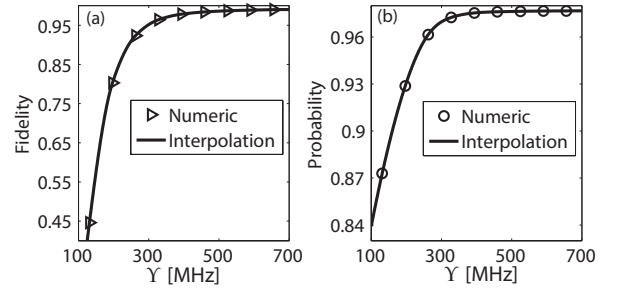,width=8cm}
      \caption{Single-photon gate performance vs. nonlinear coupling strength $\Upsilon$. Both the fidelity (a) and the probability of occupying the first Schmidt eigenmode (b) saturate as $\Upsilon$ increases. \label{fig5}}
\end{figure}
The above example creates entanglement deterministically between the signal and the pump photons. In order to implement the phase gate, however, it is necessary to ensure that these photons do not entangle at the cavity output. To this end, we propose to use photons in exponentially rising pulses whose temporal shapes replicate the ``time reversed'' cavity leakage modes \cite{kimblerev}. This allows the entire pulse of the pump photon to be in the cavity when the signal photon arrives, so that the latter's phase is switched with certainty. Using such photons, we next simulate the switching dynamics in the CQZ case, taking $Q_{\mathrm{p,f}}^{\mathrm{c}}=10^{8}$, $Q_\mathrm{s}^{\mathrm{c}}=10^{7}$, and assuming all other parameters to be the same as  in the Gaussian-pulse case considered above. The smaller $Q_{\mathrm{s}}^{\mathrm{c}}$ results in a narrower signal pulse, which allows us to arrange the temporal delay of the signal photon relative to the pump to be such that it passes through the cavity when almost the entire pump pulse is already in the cavity. The simulation results are shown in Fig.~\ref{fig4}, where plot \ref{fig4}(a) shows the signal input and output with the pump OFF. Except for the $\pi$ phase shift, the temporal profile of the signal output is reversed indicating that the signal photon coupled into the cavity and then out. Figure~\ref{fig4}(b) shows the signal input and the first Schmidt eigenfunction of the gate output with the pump ON. This first eigenfunction matches the input pulse shape very well and the fidelity reaches 0.99. The probability for the output signal photon to occupy this eigenmode is 0.98. Figure~\ref{fig4}(c) shows the pump input and output with the signal OFF. The pulse evolution is similar to the signal's in the pump-OFF case. Figure~\ref{fig4}(d) plots the pump input and the first Schmidt eigenfunction of the output with the signal ON. As shown, the pump photon couples into disk and then exits as if the signal photon did not exist. Both Figs.~\ref{fig4}(a) and \ref{fig4}(c) show that the pulses start to leak out only after they have been entirely coupled in.

We further investigate the gate performance with larger disk sizes, which lead to a range of smaller $\Upsilon$ values (cf. Fig.~\ref{fig2}). Both the fidelity and the probability of occupying the first Schmidt eigenmode are calculated and plotted in Fig.~\ref{fig5}. Both plots show a saturation feature as $\Upsilon$ increases. This feature points to the feasibility of future experiments. As long as the disk radius is smaller than \SI{25}{\micro\metre}, good single-photon switching performance can be expected.

In summary, we have proposed using the quantum Zeno effect to achieve a large effective nonlinearity at the single-photon level. By performing a multimode analysis and considering realistically achievable parameters, we have shown that a deterministic phase gate can be implemented between single photons with near-ideal fidelity.

We acknowledge D. V. Strekalov and A. S. Kowligy for helpful discussions. This research was supported in part by the National Science Foundation (Grant No. ECCS-1232022), by the Defense Advanced Research Projects Agency (DARPA) under the Zeno-based Opto-Electronics (ZOE) program (Grant No. W31P4Q-09-1-0014), and by the United States Air Force Office of Scientific Research (USAFOSR) (Grant No. FA9550-09-1-0593).

\end{document}


\title{Supplementary material for ``Photonic Nonlinearities via Quantum Zeno Blockade'' }

\author{Yu-Zhu Sun, Yu-Ping Huang, and Prem Kumar}
\date{\today}

\maketitle
\setstretch{1}
\section{Whispering Gallery Modes (WGM\lowercase{s})}

In spherical coordinates ($r$, $\theta$, $\phi$), the mode profiles for both TE and TM modes ($l$, $m$, $q$) can be written as $\Phi_{lmq}=A_{0} Y_{lm}(\theta,\phi)j_{l}(k_{lq}r)$ \cite{WGMspherical}, where $A_{0}$ is a normalization constant, $Y_{lm}$ are spherical harmonics, and $j_{l}$ are spherical Bessel functions. For brevity, we only consider the TE modes in this paper. In the large-$l$ limit, the WGM's resonance frequency varies with $l$ and $q$ as \cite{WGMspherical}:
\begin{equation}
\label{omega}
    \omega_{lq}=ck_{lq}/n=\frac{c/n}{R}\left(l+\frac{1}{2}+\alpha_{q}(\frac{l+1/2}{2})^{1/3}-\frac{n}{\sqrt{n^{2}-1}}\right),
\end{equation}
where $R$ is the disk radius, $n$ is the refractive index of the disk material, and $\alpha_{q}$ is $q$th root of the Airy function $\text{Ai}(-z)$.

\section{Calculation for $\Upsilon$ and the Data Plotted in Fig.~2}

The nonlinear coupling coefficient $\Upsilon$ is given by:
\begin{equation}
\label{upsilon}
    \begin{aligned}
        \Upsilon=\frac{\chi^{(2)}}{2}\sqrt{\frac{\hbar\omega_\mathrm{s}\omega_\mathrm{p}\omega_\mathrm{f}}{\epsilon_{0}}}\int_{V}D(\vec{r})
        \Phi_\mathrm{s}(\vec{r})\Phi_\mathrm{p}(\vec{r})\Phi_\mathrm{f}^{\ast}(\vec{r})\mathrm{d}\vec{r}.
    \end{aligned}
\end{equation}
Here $\chi^{(2)}$ is the 2nd-order nonlinearity and $D(\vec{r})$ denotes its spatial dependence as determined by the fabricated quasi-phase-matching (QPM) pattern \cite{Hae10,radialqpm}. For an azimuthally symmetric pattern \cite{Hae10} periodically poled along the cavity-field propagating direction, $D(\vec{r})$ can be expanded as \cite{Boy08}:
\begin{equation}
\label{D}
D(\vec{r})=\sum G_{j}e^{ik_{j}\phi}.
\end{equation}
Equation~(\ref{upsilon}) shows that the feasible value for $\Upsilon$ depends on both the QPM pattern and the overlap of the cavity field profiles. Since a larger $\Upsilon$ leads to better switching performance, choosing appropriate QPM pattern, disk radius, and proper WGMs for the cavity signal, pump and SF photons is important. For WGMs, $m$, $l-|m|$, and $q$ give the number of nodes along the $\phi$, $\theta$, and $r$ directions, respectively. Therefore, in order to maximize the overlap integral in Eq.~(\ref{upsilon}), we only consider the fundamental modes with $q=1$ and $l=m$. For a certain disk radius, we can find multiple resonance frequencies with different $l$'s satisfying Eq.~(\ref{omega}). To pick the right combination of modes, however, we also need to consider the constraints posed by energy conservation and the quasi-phase-matching condition:
\begin{subeqnarray}
    \label{seq1}
    \slabel{seq1a}
    \omega_{\mathrm{s}}+\omega_{\mathrm{p}}&=&\omega_{\mathrm{f}},\\
    \slabel{seq1b}
    k_{j}+m_{\mathrm{s}}+m_{\mathrm{p}}-m_{\mathrm{f}}&=&0.
\end{subeqnarray}
Here $k_{j}$ is given by $k_{j}=2\pi j/P$ with $P$ being the period of the QPM pattern. We can choose $P$ according to $m_{\mathrm{s,p,f}}$ so that $k_{1}$ satisfies Eq.~(\ref{seq1b}), and then only the first term $G_{1}e^{ik_{1}\phi}$ in Eq.~(\ref{D}) contributes to the highest value of $\Upsilon$.

With the above considerations, we numerically search for combinations of the cavity signal, pump, and SF modes within the near-infrared to the telecom-band range. For a given disk radius $R$, we are able to find 15 or more such combinations for which, $\Upsilon$ is calculated and the largest value is then plotted in Fig.~2

\section{IQZ case analysis}

\begin{figure}
 \epsfig{figure=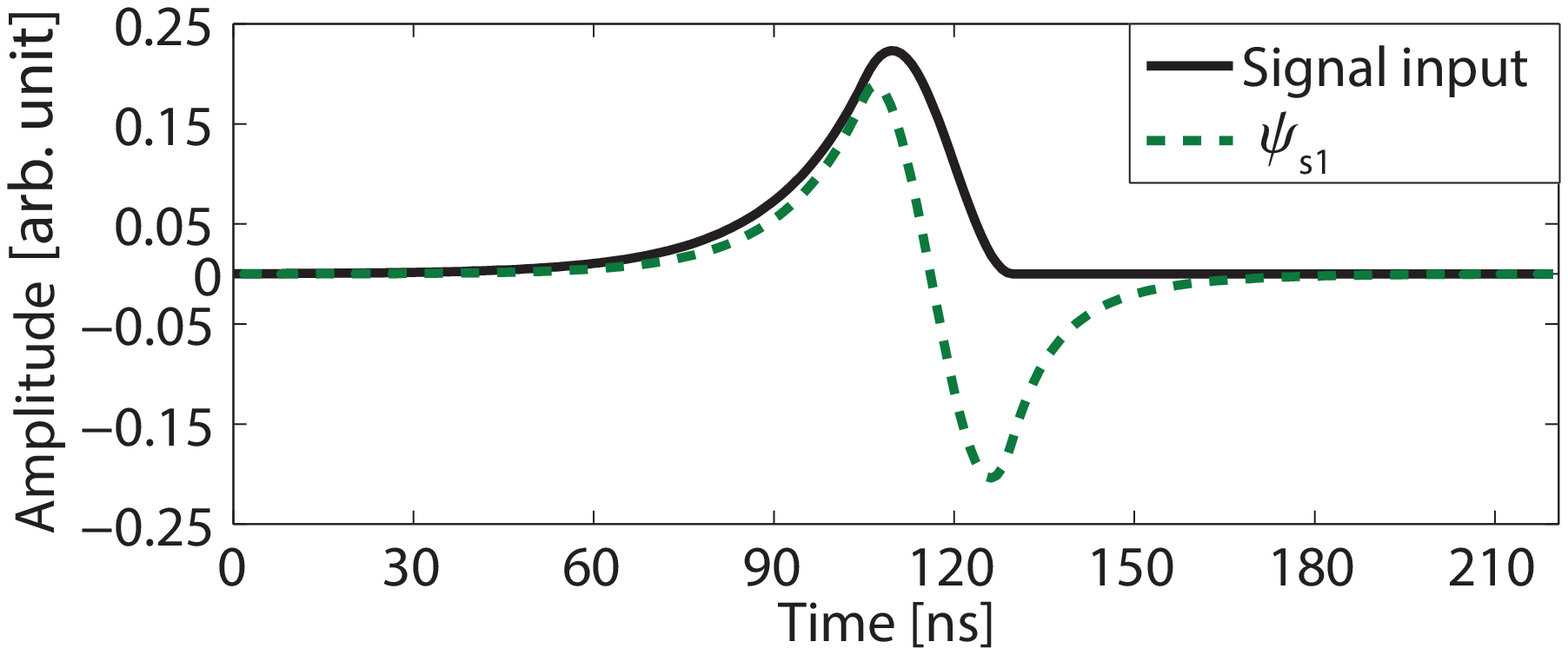,width=8cm}
      \caption{IQZ gate performance. \label{fig6}}
\end{figure}
The dissipative nature of the SF field in the cavity is incorporated by assuming a smaller $Q_\mathrm{f}^{\mathrm{i}}\sim 10^5$ while still retaining $Q_{\mathrm{s,p}}^{\mathrm{i}}\sim 10^9$. The input and the first Schmidt eigenfunction of the output for this case are shown in Fig.~\ref{fig6}. Although the probability of occupying the first eigenmode is $0.9$, the fidelity of the output with the input is only $0.51$, and the dissipation at the SF introduces $>$ $85\%$ loss to the signal energy. The performance of the switch in the IQZ case is thus significantly worse than that in the CQZ case. This is because in the presence of high loss for the cavity SF field, the efficiency for SFG is significantly reduced owing to the quantum Zeno effect. More of the signal pulse couples into the cavity, which sees loss through SFG and then leads to worse switching performance.